\documentclass[twocolumn,showpacs,preprintnumbers,amsmath,amssymb]{revtex4}
\usepackage{graphicx}
\usepackage{dcolumn}
\usepackage{bm}
\def\bea {\begin{eqnarray}}
\def\eea {\end{eqnarray}}

\def\be {\begin{equation}}
\def\ee {\end{equation}}

\begin{document}

\title{Elliptic flow of heavy flavors}

\author{Santosh K Das\footnote{santosh@vecc.gov.in} and Jan-e Alam\footnote
{jane@vecc.gov.in}}

\medskip

\affiliation{Variable Energy Cyclotron Centre, 1/AF, Bidhan Nagar , 
Kolkata - 700064}

\date{\today}
\begin{abstract}
The propagation of charm and bottom quarks through an ellipsoidal domain 
of quark gluon plasma has been studied within the ambit of non-equilibrium
statistical mechanics. Energy dissipation of heavy quarks by both radiative
and collisional processes are taken in to account. The experimental data  
on the elliptic flow of the non-photonic electrons resulting from the 
semi-leptonic decays of hadrons containing heavy flavours has been reproduced
with the same formalism that has been used earlier to 
reproduce the nuclear suppression factors.  The elliptic flow of the 
non-photonic electron from heavy meson decays produced in  nuclear collisions 
at LHC and  low energy RHIC run have also been predicted. 
\end{abstract}

\maketitle

The nuclear collisions at Relativistic Heavy Ion Collider (RHIC) 
and the Large Hadron Collider(LHC) energies 
are aimed at creating a phase where the bulk properties of the matter 
are governed by (light) quarks and gluons. Such a 
phase of matter is called Quark Gluon Plasma (QGP). 
The study of the properties of QGP is a field of high current interest.
Experimental results already available from RHIC~\cite{npa2005} have
been successfully described by the application of 
relativistic hydrodynamics~\cite{pasi,teaney} indicating substantial 
collective dynamics in the system. In
a hydrodynamic model one of the key assumption is that
the system achieves thermalization. The elliptic flow parameter
$v_2^{HF}$, is one such quantity which is sensitive to the 
early thermalization of the system. 

As the relaxation time for heavy quarks 
are larger than the corresponding quantities
for light partons~\cite{moore} 
the light quarks and the gluons get thermalized 
faster than the heavy 
quarks. Therefore, the  propagation of  heavy quarks through a 
QGP (mainly containing light quarks and gluons) 
may be treated as the 
interactions between equilibrium and non-equilibrium degrees
of freedom and the Fokker-Planck (FP) equation provides an appropriate
framework~\cite{moore,japrl,sc,svetitsky,rapp,turbide,
bjoraker,npa1997,munshi,rma} for such studies. 
Since heavy quarks remain out of equilibrium {\it i.e}
they are not a part of the equilibrated system and their production 
is restricted to the primordial stages of the collision, they can not 
decide the bulk properties of the system,  rather act as an efficient
probe to extract information about the system. Therefore, in the present 
work we will use the elliptic flow of 
heavy quarks~\cite{rappv2,gossiauxv2,hiranov2}
as a probe to extract the properties of QGP. 

In our earlier works~\cite{Das1}, we have used the FP equation 
to study the experimental results from STAR and PHENIX collaborations
on nuclear suppression factor, $R_{\mathrm AA}$
of the non-photonic electrons resulting from the semi-leptonic decays of 
hadrons containing heavy flavours~\cite{stare,phenixe}.
In the same approach  
the $R_{\mathrm AA}$'s have been predicted~\cite{Das2,Das3} for LHC and low 
energy RHIC runs~\cite{lowephenix,lowestar}.
In the present work we intend to study the elliptic flow measured
at RHIC energy with the same formalism and input parameters that 
reproduce the $R_{\mathrm AA}$. 
The $v_2^{HF}$ have 
also been predicted for low energy RHIC  and LHC energy. 

The FP equation and the procedure adopted in evaluating the drag and diffusion 
co-efficients are discussed in detail earlier~\cite{Das1,Das2,Das3}. 
Therefore, in the present work  we outline the main results and refer
to the earlier works for details. 
The evolution of heavy quarks momentum distribution function, while propagating
through the QGP are assumed to be governed by the FP 
equation which reads,
\be
\frac{\partial f}{\partial t} = 
\frac{\partial}{\partial p_i} \left[ A_i(p)f + 
\frac{\partial}{\partial p_j} \lbrack B_{ij}(p) f \rbrack\right] 
\label{expeq}
\ee
where the kernels $A_i$ and $B_{ij}$ are given by
\begin{eqnarray}
&& A_i = \int d^3 k \omega (p,k) k_i \nonumber\\
&&B_{ij} = \int d^3 k \omega (p,k) k_ik_j.
\end{eqnarray}
for $\mid\bf{p}\mid\rightarrow 0$,  $A_i\rightarrow \gamma p_i$
and $B_{ij}\rightarrow D\delta_{ij}$ where $\gamma$ and $D$ stand for
drag and diffusion co-efficients respectively.
The temperature and the baryonic chemical potential
dependence of the transport coefficients enter
through the thermal parton distributions~\cite{Das2} which
appear in $\omega(p,k)$ as
\be
\omega(p,k)=g\int\frac{d^3q}{(2\pi)^3}f^\prime(q)v\sigma_{p,q\rightarrow p-k,q+k}
\ee
The basic inputs required for solving the FP equation
are the dissipation co-efficients and initial momentum 
distributions of the heavy quarks. The drag and diffusion
coefficients have been evaluated by taking in to account
both the collisional and radiative processes~\cite{Das3}.
In the radiative process the dead cone~\cite{Kharzeev} 
and Landau-Pomerenchuk-Migdal (LPM) effects are included, the
details have been discussed in Ref.~\cite{Das3}.
In evaluating the drag and diffusion co-efficients we have
used temperature dependent  strong coupling,
$\alpha_s$ from ~\cite{Kaczmarek}.
The Debye mass, $\sim g(T)T$ is  also a temperature dependent
quantity which is used as a cut-off to shield the infrared divergences
arising due to the exchange of massless gluons.


The production of charm and
bottom quarks in hadronic collisions is studied extensively~\cite{charmbottom}.
The initial momentum distribution of heavy quarks 
in pp collisions have been taken from the
NLO MNR code~\cite{MNR}. The results  from the code
may be tested by measuring the production cross sections of
heavy mesons (containing $c$ and $b$  quarks) in pp collisions at
RHIC and LHC energies.

How can the heavy quarks
probe the asymmetry of the thermalized system formed by light quarks
and gluons? Consider a thermalized ellipsoidal spatial domain
of QGP with major and minor axes of lengths $l_y$ and  $l_x$ 
(determined by the collision geometry) respectively.
Now let us assume that a heavy quark propagates along the major
axis then the number of interactions it encounters or in other words
the amount of energy it dissipates or the amount of momentum degradation
that takes place is different from when it propagates
along the minor axis, because $l_y\neq l_x$. 
Therefore, the momentum distribution of electrons
originating from the decays of charmed hadrons ($D$ mesons) produced
from the charm quark fragmentation will be anisotropic, 
thus reflecting the spatial anisotropy due to non-central collisions
in the momentum distributions.  The degree
of momentum anisotropy will depend on both the spatial anisotropy and
more importantly on the coupling strength of the interactions 
between the heavy quark and the QGP.  
In an extreme case of zero interaction strength the 
momentum anisotropy will be zero even if the spatial anisotropy 
is large.  
For a quantitative understanding we consider the following.
The time evolution of the $i$th component 
of the average momentum for an ellipsoidal geometry 
can be obtained from the expressions: 
\be
\langle p_i\rangle\sim \exp\left(-\int^{l_i/v}\gamma(t)\,dt\right)
\ee 
where $v$ is the velocity of the heavy quark and
$l_i$ is the length along the $i$th direction.
The eccentricity of the momentum-space ellipse due to
anisotropic momentum distribution resulting from 
spatial anisotropy as discussed above is given by:
\be
\epsilon_p=\frac{\langle p_x^2\rangle-\langle p_y^2\rangle}
{\langle p_x^2\rangle+\langle p_y^2\rangle}
\label{epsilonp}
\ee
where $\epsilon_p$ is the eccentricity in the momentum space.
The values of $l_y$ and $l_x$ are fixed by geometry of
the collisions and can be estimated using 
Glauber model for different centralities. 
Both $l_x\neq l_y$ and $\gamma\neq 0$ (non-zero interaction)
are required for non-zero elliptic flow.
The second Fourier coefficient of the
momentum distributions is called elliptic flow, $v_2^{HF}$ and 
we evaluate this quantity in the present work. 


The FP equation has been solved for the heavy quarks by using the Greens 
function technique (see~\cite{rapp} for details). 
The initial $p_T$ distribution is taken from NLO MNR formalism~\cite{MNR}.
We convolute the solution (at the transition point
{\it i.e.} when the QGP reverts back to hadronic phase)
with the Peterson fragmentation function~\cite{peterson} of the
heavy quarks to obtain the $p_T$ distribution of the
heavy ($B$ and $D$) mesons. 
The momentum distribution, $dN/dydp_Td\phi$ of the 
non-photonic single electron spectra originating from the 
decays of heavy flavour mesons - {\it e.g.} $D\rightarrow Xe\nu$ 
at mid-rapidity ($y=0$) is obtained by following the
procedure described in ~\cite{Das1,gronau,ali}. The coefficient of elliptic 
flow, $v_2^{HF}$ then can be obtained as:

\be
v_2^{HF}(p_T)=\langle cos(2\phi) \rangle= \frac{\int d\phi \frac{dN}{dydp_Td\phi×}|_{y=0} cos(2\phi)}
{\int d\phi \frac{dN}{dydp_Td\phi×}|_{y=0}×}
\ee

The system formed in nuclear collisions at relativistic energies
evolves dynamically from the initial to the final state. 
The time evolution of this systems has been studied 
by using the (2+1) dimensional hydrodynamical model~\cite{hydro}
with boost invariance along the longitudinal direction~\cite{bjorken}.
It is expected that the central rapidity
region of the system formed after nuclear collisions
at LHC energy is almost net baryon free. Therefore,
the equation governing the conservation of net
baryon number need not be considered for RHIC and LHC`. 

The total amount of energy dissipated by a  heavy quark in the QGP
depends on the path length it traverses.
Each parton traverse different path length
which depends on the  geometry of the system and on the point 
where its is created.
The probability that a parton is produced at a point $(r,\phi^\prime)$
in a QGP of ellipsoidal shape depends on the number of binary collisions 
at that point which can be taken as:
\be
P(r,\phi^\prime)=\frac{1}{{\cal N}}\left(1-\frac{r^2}{R^2}
\frac{(1+\epsilon cos^2\phi^\prime)}{(1-\epsilon^2)^2}\right)\Theta(R-r)
\label{prob}
\ee
and
\be
{\cal{N}}=\frac{1}{\pi R^2\left(1-\frac{1}{2}\frac{1+\epsilon^2/2}
{(1-\epsilon^2)^2}\right)
}
\ee
where $R$ is the nuclear radius 
and $\epsilon$ is the eccentricity of the ellipse. 
Note that for central collisions ($\epsilon=0$) Eq.~\ref{prob}
reduces to the expression for $P(r,\phi^\prime)$ used in Ref.~\cite{turbide}
for spherical geometry. 
A parton created at $(r,\phi^\prime)$ in the transverse plane
propagate a distance $L=\sqrt{R^2-r^2sin^2\phi^\prime}-rcos\phi^\prime$
in the medium. We use the following
equation for the geometric average of the integral which appear
in the solution of the FP equation
involving drag coefficient: 
\be
\Gamma=\frac{\int rdr d\phi^\prime P(r,\phi^\prime) \int^{L/v}d\tau\gamma(\tau)}
{\int rdr d\phi^\prime P(r,\phi^\prime)}
\label{cgama}
\ee
where $v$ is the velocity of the propagating partons. 
Similar averaging has been performed  for the diffusion co-efficient.
For a static system the temperature dependence of the drag and
diffusion co-efficients of the heavy quarks enter via the
thermal distributions of light quarks and gluons through
which it is propagating. However, in the present scenario
the variation of temperature with time is governed by
the equation of state or the velocity of sound
of the thermalized system undergoing hydrodynamic
expansion. In such a scenario the quantities like $\Gamma$ (Eq.~\ref{cgama})
and hence $v_2$ becomes sensitive to velocity of sound ($c_s$)
in the medium. 

\begin{figure}[h]
\begin{center}
\includegraphics[scale=0.4]{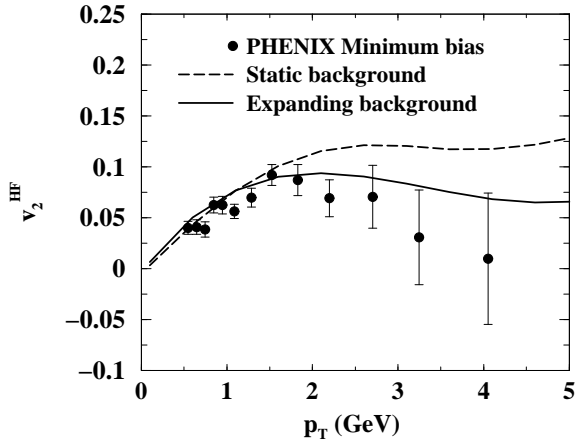}
\caption{Variation of $v_2^{HF}$ with $p_T$ at  the
highest RHIC energy. Experimental data is taken from~\cite{phenixemb}.}
\label{fig1}
\end{center}
\end{figure}

\begin{figure}[h]
\begin{center}
\includegraphics[scale=0.4]{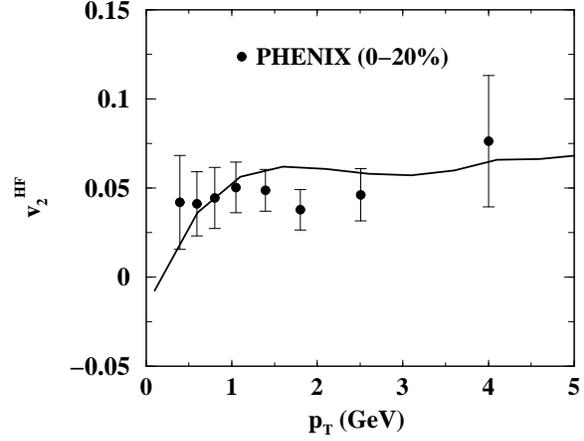}
\caption{Same as Fig.~\ref{fig1} for 0-20\% centrality.
Experimental data contain both the statistical and 
systematic errors,  taken from~\cite{phenixecen}. 
 }
\label{fig2}
\end{center}
\end{figure}

\begin{figure}[h]
\begin{center}
\includegraphics[scale=0.4]{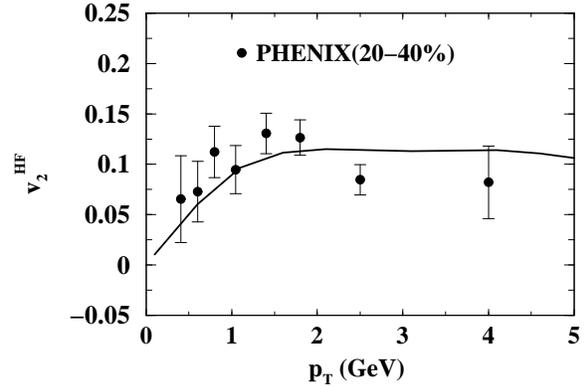}
\caption{Same as Fig.~\ref{fig1} for 20-40\% centrality.
Experimental data contain both the statistical  as well as
systematic errors,  taken from~\cite{phenixecen}. 
}
\label{fig3}
\end{center}
\end{figure}

\begin{figure}
\begin{center}
\includegraphics[scale=0.4]{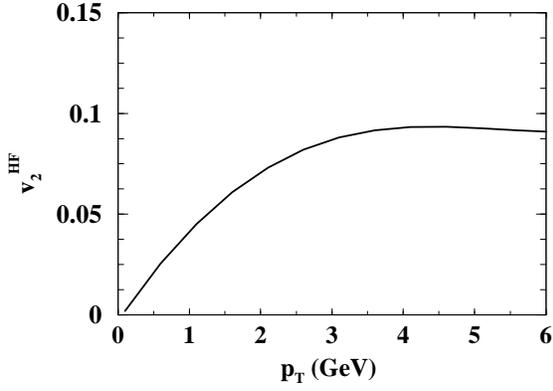}
\caption{The variation of $v_2^{HF}$ with $p_T$ for LHC initial conditions
for 0-10\% centrality.}
\label{fig4}
\end{center}
\end{figure}

\begin{figure}[h]
\begin{center}
\includegraphics[scale=0.4]{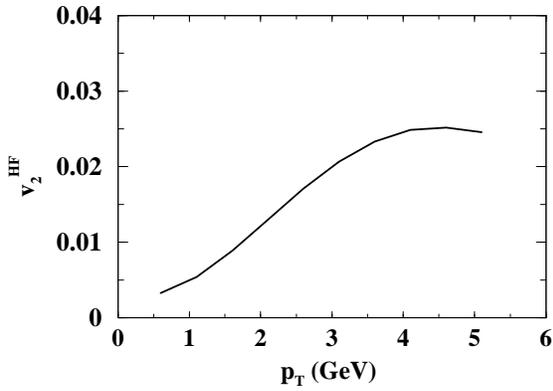}
\caption{The variation of $v_2^{HF}$ with $p_T$ for low energy RHIC
initial conditions for 0-10\% centrality.
}
\label{fig5}
\end{center}
\end{figure}


The $p_T$ dependence of $v_2^{HF}$ is sensitive to the nature of the initial 
distributions of heavy quarks, life time of the system, velocity of sound as well as the drag and 
diffusion coefficients of heavy quarks.

The experimental
data ~\cite{phenixemb} at RHIC energy ($\sqrt{s_{NN}}$=200 GeV)
shows non-zero elliptic flow indicating
substantial interactions of the plasma particles with charm and bottom quarks
from which electrons are originated through the process:
$c(b)$ (hadronization)${\longrightarrow}$ $D(B)$(decay)$\longrightarrow$
$e+X$. 
In Fig.~\ref{fig1} we compare the experimental data obtained by  the
PHENIX~\cite{phenixemb} collaborations 
for Au + Au minimum bias collisions at $\sqrt{s_{\mathrm NN}}=200$ GeV with
theoretical results obtain in the present work. We observe that the 
data can be reproduced by considering both radiative and collisional loss
with $c_s=1/\sqrt{4}$. 
In this case $v_2^{HF}$ first increases and reaches a maximum of about 8\% 
then  saturates for  $p_T>2$ GeV. From the energy dissipation
we have evaluated the shear viscosity ($\eta$) to entropy ($s$) density ratio
using the relation~\cite{mmw}: $\eta/s\sim 1.25T^3/\hat{q}$, where $\hat{q}=
\langle p^2_T\rangle/L$ and $dE/dx\sim \alpha_s\langle p^2_T\rangle$
~\cite{RB}, $L$ is the length traversed by the heavy quark  and $p_T$
is its transverse momentum. The average value of $\eta/s\sim 0.1-0.2$,
slightly above the AdS/CFT bound~\cite{KSS}. 
In Fig.~\ref{fig1} we also display the results obtained when the expansion
of the background matter is switched off (dashed line). In this case
values of  the drag and diffusion co-efficients are evaluated
at the initial temperature. The magnitude of 
$v_2$ is more when background is static, the reason for this can be understood
from the following. 
The $v_2$  controls the length of the major axis $(=1+v_2)$
and minor  axis $(=1-v_2)$ and therefore, it also determine the 
value of the eccentricity as follows:
\be
\epsilon_p=\sqrt{1-\frac{(1-v_2)^2} 
{(1+v_2)^2}} 
\label{epsilonv2}
\ee
Eliminating $\epsilon_p$ from Eqs.~\ref{epsilonp} and ~\ref{epsilonv2}
we can write 
\be
v_2=\left[\frac{\sqrt{\langle p_x^2\rangle}-\sqrt{\langle p_y^2\rangle}}
{\sqrt{\langle p_x^2\rangle}+\sqrt{\langle p_y^2\rangle}}\right]^2
\label{v2static}
\ee
Absence of expansion of the background matter is unable to 
smoothed out the prevailing momentum anisotropy ($\langle p_x^2\rangle
\neq \langle p_y^2\rangle$) in the
system  hence a larger value of $v_2$ is obtained for
a static system as indicated by Eq.~\ref{v2static}. The 
difference between the dashed and solid lines in Fig.~\ref{fig1}
also indicates the role of the expansion of the background vis-a-vis
initial geometric anisotropy. We would like to mention that
except the results indicated by dashed line in Fig.~\ref{fig1}
all other results are obtained with the background expansion
within the ambit of (2+1) dimensional hydrodynamics.

The comparison of results obtained in the present 
work with the experimental data ~\cite{phenixecen} obtained
for two different centralities are depicted 
in Figs.~\ref{fig2} and ~\ref{fig3}. The agreement is 
reasonable both for $0-20\%$ and $20-40\%$ centralities.

The $v_2^{HF}$ (for 0-10\% centrality collision) at LHC energy 
has been obtained by using the current formalism is
displayed in Fig.~\ref{fig4}. The variation of $v_2^{HF}$ 
with $p_T$ is similar to that of RHIC.


Now we turn to collisions at low energy, $\sqrt{s_{\mathrm NN}}=39$ GeV. 
For the initial conditions the $p_T$ distribution of 
heavy quarks in $pp$ collisions is required.
At low collision energy rigorous QCD based calculations for 
heavy flavour production is not available. 
In the present work 
this is obtained from pQCD calculation~\cite{pqcd,combridge}
for the processes: $gg\rightarrow Q\bar{Q}$
and $q\bar{q}\rightarrow Q\bar{Q}$.  
The composition of the matter formed at low energy 
collisions will be different from that of the matter
formed at highest RHIC/LHC energies. The net baryon
number will
be non-zero for low energy collisions. Therefore, in addition
to temperature we need another thermodynamic variable, the
baryonic chemical potential, $\mu_B$ to characterize the matter.
As a result in addition to energy momentum conservation
equation which is used to study the evolution of baryon free matter 
we need to solve the baryon number conservation equation also.
More details about the initial conditions,
$T$ and $\mu_B$ dependence of the transport coefficients
and the space time evolution have been outlined in~\cite{Das2}.

At low $\sqrt{s_{\mathrm NN}}$(=39 GeV) the net baryon
density at mid-rapidity is non-zero and its value
could be high depending on the value of   
 $\sqrt{s_{\mathrm NN}}$. Therefore, we need to solve
the FP equation for both non-zero $T$ and $\mu_B$ which enters
the equation through the drag and diffusion
coefficients.
The baryonic chemical potential for  
$\sqrt{s_{\mathrm NN}}$=39 GeV 
has been  obtained from the parametrization of the  experimental data on
hadronic ratios as~\cite{ristea} (see also~\cite{andronic}),
\be
\mu_B(s_{\mathrm NN})=a(1+\sqrt{s_{\mathrm NN}}/b)^{-1}
\label{mub}
\ee
where $a = 0.967\pm 0.032$ GeV and $b=6.138 \pm 0.399$ GeV.
The parametrization in Eq.~\ref{mub} gives the values
of $\mu_B$ at the freeze-out. The corresponding value
at the initial condition is obtained from the 
baryon number conservation equation. 

In Fig.~\ref{fig5}, $v_2^{HF}$ for $\sqrt{s_{\mathrm NN}}=39$
GeV has been displayed. The lack 
of saturation in the trend of $v_2^{HF}$ reflect insignificant B-meson 
contribution as well as the short life time of the plasma. 

So far we have discussed the azimuthal asymmetries of the
non-photonic electron produced in nuclear collisions
due to the propagation of the heavy quark in the
partonic medium in the pre-hadronization era. However,
the azimuthal asymmetries of the D mesons in the post
hadronization era (when both the temperature and
density are lower than the partonic phase) should in
principle be also taken into account. The suppression of the D 
mesons in the post hadronization era is found  to be negligibly
small ~\cite{Das2}, indicating the fact that
the hadronic medium (of pions and nucleons) is unable
to drag the D mesons. In the same spirit we have neglected 
the azimuthal asymmetries due to the hadronic interactions.

In summary, we have evaluated the azimuthal asymmetries
($v_2^{HF}$) using the Fokker-Planck  equation
with the same set of inputs that reproduces the nuclear suppression,
$R_{AA}$ measured experimentally at the highest RHIC energy. 
We found that the $B$ meson suffer less flow than $D$ mesons.
The B-meson contribution reflects itself by the saturation of $v_2^{HF}$ 
in both LHC and RHIC energy and lack of this trend in low energy RHIC 
run reflect the insignificant contribution of B-meson. The 
flow pattern at LHC is similar  to that of RHIC.
Some comments on the $R_{\mathrm AA}$  vis-a-vis $v_2$
flow are in order here.  The 
$R_{\mathrm AA}$  contains the ratio of $p_T$ distribution 
of electron resulting from Au+Au to p+p collisions, where
the numerator contains the interaction of the heavy quarks
with the medium (QGP) and such interactions are absence in 
the denominator. Whereas for $v_2$ both the numerator and the
denominator contain the interactions with the medium, resulting in
some sort of cancellation. Therefore, the $R_{AA}$ will be
sensitive to the $K$ factor but one should not expect the same
kind of sensitivity in $v_2$. 

{\bf Acknowledgment:}
This work is supported by DAE-BRNS project Sanction No. 2005/21/5-BRNS/2455.
We are grateful to Matteo Cacciari for providing us the heavy
quarks transverse momentum distribution for pp collisions.  
We are also thankful to Victor Roy for useful discussions on (2+1)
dimensional hydrodynamics.


\begin{thebibliography}{99}

\bibitem{npa2005} I. Arsene {\it et al.} (BRAHMS Collaboration),
Nucl. Phys. A {\bf 757}, 1 (2005); B. B. Back {\it et al.} (PHOBOS 
Collaboration), Nucl. Phys. A {\bf 757}, 28 (2005);
J. Adams {\it et al.} (STAR Collaboration), Nucl. Phys. A {\bf 757}, 102 (2005);
K. Adcox {\it et al.} (PHENIX Collaboration), 
Nucl. Phys. A {\bf 757}, 184,(2005).

\bibitem{pasi} P. Huovinen and P. V. Ruuskanen, 
Ann. Rev. Nucl. Part. Sci. {\bf 56}, 163 (2006).

\bibitem{teaney} D. A. Teaney, arXiv:0905.2433 [nucl-th].

\bibitem{moore} G. D. Moore and D. Teaney, Phys. Rev. C {\bf 71}, 064904
 (2005).

\bibitem{japrl} J. Alam, S. Raha and B. Sinha, Phys. Rev. Lett. {\bf 73}, 1895
(1994).

\bibitem{sc} S. Chakraborty and D. Syam, Lett. Nuovo Cim. {\bf 41}, 381 (1984).

\bibitem{svetitsky} B. Svetitsky, Phys. Rev. D {\bf 37}, 2484( 1988).

\bibitem{rapp} H. van Hees, R. Rapp, Phys. Rev. C,{\bf 71}, 034907 (2005).

\bibitem{turbide} S. Turbide, C. Gale, S. Jeon and G. D. Moore,
Phys. Rev. C {\bf 72}, 014906 (2005).

\bibitem{bjoraker} J. Bjoraker and R. Venugopalan, Phys. Rev. C {\bf 63},
024609 (2001).

\bibitem{npa1997} P. Roy, J. Alam, S. Sarkar, B. Sinha and S. Raha,
Nucl. Phys. A {\bf 624}, 687 (1997)

\bibitem{munshi} M  G. Mustafa and  M. H. Thoma, Acta Phys. Hung. A
{\bf 22}, 93 (2005).

\bibitem{rma} P. Roy, A. K. Dutt-Mazumder and J. Alam, Phys. Rev. C {\bf 73},
044911 (2006).

\bibitem{rappv2} H. van Hees, V. Greco, R. Rapp, Phys. Rev. C,{\bf 73}, 034913 (2006).

\bibitem{gossiauxv2}
P. B. Gossiaux, and J. Aichelin, Phys. Rev. C,{\bf 78}, 014904 (2008)


\bibitem{hiranov2} Y. Akamatsu, T. Hatsuda and T. Hirani, 
Phys. Rev. C {\bf 79}, 054907 (2009)

\bibitem{Das1}S. K Das, J. Alam and P. Mohanty, 
Phys. Rev. C {\bf 80}, 054916 (2009)

\bibitem{stare} B. I. Abeleb {\it et al.} (STAR Collaboration), Phys. Rev.
Lett. {\bf 98}, 192301 (2007).

\bibitem{phenixe} S. S. Adler {\it et al.} (PHENIX Collaboration),
Phys. Rev. Lett. {\bf 96}, 032301 (2006).

\bibitem{Das2}S. K Das, J. Alam, P. Mohanty and B. Sinha
Phys. Rev. C {\bf 81}, 044912 (2010).

\bibitem{Das3}S. K Das, J. Alam and P. Mohanty, 
Phys. Rev. C {\bf 82}, 014908 (2010).

\bibitem{lowephenix} T. Sakaguchi (PHENIX collaboration), 
arXiv:0908.3655 [hep-ex].

\bibitem{lowestar} B. I. Abelev (Star Collaboration),
arXiv:0909.4131 [nucl-ex]

\bibitem{Kharzeev} Y. L. Dokshitzer and D. E. Kharzeev, Phys. Lett. B, 
{\bf 519}, 199 (2001).

\bibitem{Kaczmarek} O. Kaczmarek and F. Zantow, 
Phys. Rev. D, {\bf 71}, 114510 (2005).

\bibitem{charmbottom} M. Cacciari, S. Frixione,  M.L. Mangano,
P. Nason and  G. Ridolfi, J. High Ener. Phys.  {\bf 0407}, 033 (2004);
M. Cacciari and P. Nason, Phys. Rev. Lett. {\bf 89}, 122003 (2002);
M. Cacciari and  P. Nason, J. High Ener. Phys.  {\bf 0309}, 006 (2003);
M. Cacciari, P. Nason and R. Vogt, Phys. Rev. Lett.
{\bf 95}, 122001 (2005).


\bibitem{MNR} M. L. Mangano, P. Nason and G. Ridolfi, Nucl. Phys.
B {\bf 538}, 282 (2002).

\bibitem{peterson} C. Peterson {\it et al.},  Phys. Rev. D {\bf 27}, 105 (1983).

\bibitem{gronau} M. Gronau, C. H. Llewellyn Smith,
T. F. Walsh, S. Wolfram and T. C. Yang,
Nucl. Phys. B {\bf 123}, 47 (1977).

\bibitem{ali} A. Ali, Z. Phys. C {\bf 1}, 25 (1979)

\bibitem{hydro} P. F. Kolb, J. Sollfrank and U. Heinz, Phys. Rev. C {\bf 62}, 054909
(2000); P. F. Kolb and R. Rapp, Phys. Rev. C {\bf 67}, 044903 (2003);
P. F. Kolb and U. Heinz, nucl-th/0305084,
J. Sollfrank, P. Koch and U. Heinz, Phys. Lett. B {\bf 252}, 256 (1990)
J. Sollfrank, P. Koch and U. Heinz, Z. Phys. C {\bf 52}, 593 (1991).

\bibitem{bjorken} J. D. Bjorken, Phys. Rev. D {\bf 27}, 140 (1983).


\bibitem{phenixemb} S. S. Adler {\it et al.} (PHENIX Collaboration),
Phys. Rev. Lett. {\bf 98}, 172301 (2007)

\bibitem{mmw} A. Majumder, B. M\"uller and X. N. Wang, Phys. Rev. Lett.
{\bf 99}, 192301 (2007).

\bibitem{RB} R. Baier, arXiv hep-ph/0209038.

\bibitem{KSS} P. Kovtun, D. T. Son and A. O. Starinets, Phys. Rev. Lett. 
{\bf 94}, 111601 (2005).

\bibitem{phenixecen}A. Adare {\it et al.} (PHENIX Collaboration), arXiv 
1005.1627[nucl-ex].

\bibitem{pqcd} R. D. Field, Application of Perturbative QCD, Addison-Wesley
Pub. Company, N.Y. 1989.

\bibitem{combridge} B.L. Combridge, Nucl.Phys.B {\bf 151}, 429 (1979).

\bibitem{ristea} O. Ristea (for the BRAHMS collaboration)
Romanian Reports in Physics, {\bf 56}, 659(2004)

\bibitem{andronic} A. Andronic, P. Braun-Munzinger and J. Stachel, 
Nucl. Phys. A {\bf 772}, 167 (2006).  

\end{thebibliography}
\end{document}